\documentclass[prl,aps,floatfix,twocolumn,superscriptaddress]{revtex4}

\usepackage{amsmath}
\usepackage{graphicx}

\begin{document}

\title{THz emission from ultrafast optical orientation}
\author{F. Nastos}
\affiliation{Department of Physics and Institute for Optical
Sciences, University of Toronto, 60 St. George Street, Toronto,
Ontario, Canada M5S 1A7}
\author{R. W. Newson}
\affiliation{Department of Physics and Institute for Optical
Sciences, University of Toronto, 60 St. George Street, Toronto,
Ontario, Canada M5S 1A7}
\author{J. H{\"u}bner}
\affiliation{Institute for Solid State Physics, University of
Hannover, Appelstr. 2, 30167 Hannover, Hannover, Germany}
\author{J. E. Sipe}
\affiliation{Department of Physics and Institute for Optical
Sciences, University of Toronto, 60 St. George Street, Toronto,
Ontario, Canada M5S 1A7}
\author{H. M. van Driel}
\affiliation{Department of Physics and Institute for Optical
Sciences, University of Toronto, 60 St. George Street, Toronto,
Ontario, Canada M5S 1A7}

\begin{abstract}
We show both theoretically and experimentally that the magnetization
density accompanying ultrafast excitation of a semiconductor with
circular polarized light varies rapidly enough to produce a
detectable THz field.
\end{abstract}

\maketitle

A host of recent fundamental studies on the behavior of spins in
semiconductors have been motivated by the idea of using the spin
degree of freedom in information processing, the goal of a field
that has come to be known as \textit{spintronics}.  Central to many
of these investigations has been the optical injection of spin
polarized electrons and holes into the band structure of a
semiconductor.  These spin polarized carriers are then available to
be dragged by a bias voltage, for example, to regions of interest in
any proposed device structure.  Such optical injection of spin
polarized carriers, also called \textit{optical orientation}, can be
effected by the absorption of circularly polarized light
\cite{DyakanovPerel,ZuticRMP04}, which takes advantage of the
selection rules to populate particular angular momentum states of
the band structure.

In the analysis of most of these experiments it is assumed that the
holes can be neglected, because in bulk semiconductors the hole spin
lifetimes are typically much shorter than the electron spin
lifetimes. This makes it difficult, if not impossible, to measure
the optical orientation of hole spins in bulk semiconductors using
usual techniques employed to study electron spin orientation.
However, Hilton and Tang \cite{HiltonPRL02} recently showed that
pump-probe techniques could be used to probe the hole spin dynamics
in bulk GaAs.  They found a time of $\tau_h = 110\,\text{fs}$ for
the hole spin relaxation time, which has spurred theoretical
investigations \cite{YuPRB05}.  All such techniques measure optical
orientation {\em indirectly} by monitoring, after excitation, the
population in various bands, or its effect on luminescence or
Faraday rotation \cite{OestrichSST17,AwschalomPT52}.

In this manuscript we present an approach for the \textit{direct}
measurement and study of ultrafast optical orientation.  While we
focus on GaAs, the technique is applicable to a wide range of
semiconductors.  It relies on the magnetization that accompanies the
injected spin density in optical orientation: Under ultrafast
excitation this magnetization varies rapidly over a sub-picosecond
timescale, and radiates in the terahertz (THz) regime.  Such
radiation has already been used to study a host of other ultrafast
processes
\cite{BeaurepaireAPL04,SchmuttenmaerCR04,JohnstonPRB02,CoteAPL02},
and we show here that detailed measurement and analysis of the
emitted THz electric field trace can be used to study the electron
and hole spin dynamics.

To estimate the THz field strength radiated from ultrafast optical
orientation, we compare the injected magnetization source $M(t)$ in
optical orientation to the injected polarization source $P(t)$ in
the shift current (also called the intrinsic photovoltaic effect
\cite{vonBaltzPRB81}).  When a semiconductor is excited by photons
with energies above the band gap, electrons and holes are injected
at a rate that closely follows the temporal pulse profile
$\mathcal{I}(\mathbf{r},t)$, which is the intensity profile of the
laser pulse in the sample.  The carrier density injection rate
$\dot{n}$ is given by $\dot{n}(\mathbf{r},t)=\xi_\omega
\mathcal{I}(\mathbf{r},t)$, where $\omega$ refers to the carrier
frequency of the laser, and $\xi_\omega$ is proportional to the
absorption coefficient. The spatial coordinate $\mathbf{r}$ accounts
for the variation from the exponential decay into the surface, and
the variation in intensity across the spot-size.  The polarization
and magnetization sources we consider both arise from the carrier
injection, although they have different symmetry characteristics and
underlying microscopic origins.

In III-V semiconductors, the shift current appears for excitation by
certain {\em linear} polarizations of the field
\cite{noteAboutPGEffect}.  In this effect the center of charge of
the carriers moves as the carriers are promoted from the valence
band to the conduction band, and a net current directly results,
\begin{equation}
\dot{P}_{\text{inj}}(t) = p\dot{n}(\mathbf{r},t), \label{pdot}
\end{equation}
where $p=ed$ is the average injected dipole moment per carrier.
Here, $e$ is the electron charge and the displacement $d$ is on the
order of a Bohr radius $a_\text{B}$.  Detailed calculations show
that for GaAs under excitation with linear polarized light along
[110] this distance $d$ is very close to the GaAs bond length,
$d=2.54\,a_\text{B}$ \cite{NastosPRB06}. Since the polarization
$P_{\text{inj}}(t)$ rises on a subpicosecond time-scale, it radiates
in the THz regime \cite{neglectOR}. However, the polarization
survives for much longer than a picosecond, and so its decay is not
expected to radiate in the THz regime. Thus for the polarization
that contributes to the far field radiation we have $\ddot{P}(t) =
\ddot{P}_{\text{inj}}(t)$.

Optical orientation arises from {\em circularly} polarized
excitation of carriers near the band edge.  Because of spin-orbit
splitting these carriers can have a net spin polarization, and
accompanying this spin density is a magnetization density
$M(\mathbf{r},t)$.  For optically injected electrons, close to the
conduction band edge, the injected magnetization density is given by
\begin{equation}
\dot{M}_{e,\text{inj}}(\mathbf{r},t) = \bar\mu_e
\dot{n}(\mathbf{r},t), \label{mdot}
\end{equation}
where $\bar{\mu}_e$ is the mean magnetic moment of an injected
electron.  Unlike for free electrons -- where the magnetic moment of
an electron is simply $\mu = g\mu_{\text B} s/\hbar$, where
$\mu_{\text B}$ is the Bohr magneton, $s=\hbar/2$, and $g$ is the
free electron $g$-factor, $g=2.0023$  -- the average magnetic moment
of an injected electrons can be written as $\bar\mu_e = g_e^\ast
\mu_{\text B} S / \hbar$, where $g_e^\ast$ is an effective
$g$-factor of an electron, and $S$ is the injected degree of spin
polarization in units of $\hbar/2$.  The effective $g$-factor can be
calculated from Roth's expression for the magnetic moment of a Bloch
state in the $n$-th band~\cite{RothPR60},
\begin{equation}
\mu_n^z = - \frac{g \mu_\text{B} \sigma^z_n}{2}
-\frac{2}{mi}\sum_{s\ne n}
\frac{p^x_{ns}p^y_{sn}-p^y_{ns}p^x_{sn}}{\epsilon_{n}-\epsilon_{s}},
\label{Roth}
\end{equation}
and defining the effective $g$-factor {\it via} $g_e^\ast =
-2\mu_n/\mu_\text{B}$.  In Eq.~\eqref{Roth}, $\sigma^z_n$ is the
expectation value of the Pauli spin matrix in the $n$-th band,
$p^x_{ns}$ is the $x$ component of the momentum matrix element
between the bands $n$ and $s$, and $\epsilon_n$ is the energy of the
$n$-th band.  The $z$ direction is the direction along which the
magnetization is aligned. The appearance of $\sigma^z_n$ in the
first term is required to generalize the more common $g$-factor
expression \cite{ChadiPRB76} beyond $j=1/2$ states.  For optically
injected electrons in GaAs, close to the conduction band edge,
$g_e^\ast$ is $-0.44$ at low temperatures~\cite{HermannPRB77}.

This magnetization density $M_e(t)$ will vary on an ultrafast
time-scale with the laser pulse, and it will radiate in the THz
regime.  The radiation will be magnetic-dipole like, in contrast to
the electric-dipole like radiation from the shift current.  Since
the lifetime of the injected spins is known to be on the order of
many picoseconds~\cite{ZuticRMP04}, their decay will not contribute
to the THz and we have
$\ddot{M}_e(\mathbf{r},t)=\ddot{M}_{e,\text{inj}}(\mathbf{r},t).$

The far-field radiation from both magnetic and polarization sources
is dependent on the second derivatives of the sources, $\ddot{M}(t)$
and $\ddot{P}(t)$.  A simple estimate of the radiation from just the
electron magnetization ${M}_{e}(t)$ can be made by comparing this
source to the polarization induced by the shift current.  Since both
of these sources survive long enough that their decay profiles are
not expected to radiate in the THz regime, it is mainly the
generation (or turn on) of these sources that provides the THz
radiation.  The ratio of the peak strengths for the magnetization
and polarization sources are directly comparable, since the
different Fresnel coefficients governing the transmission of
radiation for each type of source only determine the radiation
patterns.  So to first order we can estimate the strength of the
magnetic radiation from the ratio $\ddot{M}_e(t) / \ddot{P}(t)$.
Using Eq.~(\ref{mdot}) and~(\ref{pdot}) gives
\begin{equation}
\frac{\ddot{M}_e}{\ddot{P}}= \frac{\bar\mu_e}{p} =
\frac{Sg_e^\ast\mu_{\text B}}{\hbar ed}=
\frac{(S/\frac{\hbar}{2})g_e^\ast}{4(d/a_\text{B})}\alpha,
\label{eq1}
\end{equation}
where $\alpha$ is the fine structure constant.  In GaAs optical
orientation gives 50\% spin polarization ($S=0.5 \hbar/2$), so the
ratio $\ddot M_e/\ddot P$ is roughly $1.6\times 10^{-4}$.   The THz
fields from the shift current have been measured to be about 1500
V/m, so an absolute sensitivity of about 0.2 V/m is needed to
measure the THz radiation from optical orientation of bulk GaAs
electrons alone.  This requirement is experimentally challenging
\cite{LloydHughesPRB04} but would not be impossible.  Note that this
is because the THz field amplitude, rather than the intensity, is
measured experimentally.  Were the intensity measured, the signal
from the magnetization effect would be down from that of the shift
current by a factor of $(\ddot M_e/\ddot P)^{2}$.

The relatively short lifetime of the holes, however, suggests that
the hole spins may radiate in the THz regime.  Including the decay
of the hole spins, the magnetization of the holes follows
\begin{equation}
\dot{M}_h(\mathbf{r},t) = \dot{M}_{h,\text{inj}}(\mathbf{r},t) -
\frac{{M}_h(\mathbf{r},t)}{\tau_h},
\end{equation}
where $\dot{M}_{h,\text{inj}}$ is the hole injection rate
\begin{equation}
\dot{M}_{h,\text{inj}}(\mathbf{r},t)=\bar\mu_h\dot{n}(\mathbf{r},t),
\end{equation}
analogous to the electron injection rate, but here $\bar\mu_h$ is
the mean magnetic moment of the injected holes.

On the one hand, the decay of the holes adds more temporal variation
to the magnetization, and enhances the THz signal;  on the other
hand, since the decay lowers the overall magnitude of the
magnetization, it is expected to decrease the THz signal. To include
both effects, we have solved for the far-field radiation from an
injected magnetization using a Green function formalism tailored to
planar interfaces \cite{SipeJOSAB87}.  We find that for pulse widths
of 100\,fs the effect of the hole spin decay
($\tau_h=110\,\text{fs}$) is to decrease the peak THz radiation from
this source by roughly a factor of two.  That is, the hole spins
decay fast enough so that the decay process will contribute the THz
radiation, but slow enough that the overall signal from it is not
washed out.

Yet the relatively short lifetimes of holes, in bulk materials such
as GaAs, prevents the study of their dynamics by many other
techniques. Thus the hole $g$-factors, and their magnetic moments in
general, have been largely neglected in bulk materials.  In GaAs
based nanostructures however, hole $g$-factors have been the subject
of much recent attention \cite{PryorPRL06}. It is typically found
that in those systems that the holes have a $g$-factor many times
larger than the electron $g$-factor.  In this work however, we are
interested in the optically injected magnetic moments $\bar\mu_h$,
rather than the $g$-factors themselves.  The average magnetic moment
$\bar\mu_h$ will depend on the details of the populations injected
into the heavy-hole and light-hole bands. We calculate directly the
injected hole magnetization into these states, but for easy
comparison to the electron magnetization, we quote our result as a
hole $g$-factor $\tilde{g}_h^\ast$ relative to the injected electron
spin density, so that
\begin{equation}
\bar\mu_h = \frac{ \tilde{g}_h^{\ast} \mu_{\text B} S}{\hbar}.
\end{equation}
The tilde is to emphasize that we are not defining the $g$-factor
relative to the hole spins.  Following a similar approach as
in~\cite{BhatPRL05} we use Fermi's golden rule together with Roth's
formula Eq.~\eqref{Roth} for the magnetization, and use a 30-band
$\mathbf{k}\cdot\mathbf{p}$ band structure~\cite{RichardPRB04} to
evaluate $\dot{M}_{h,\text{inj}}$ directly.  We find that
$\tilde{g}^{\ast}_h = 14.1$, which implies that the hole
magnetization injection is approximately 32 times larger than the
electron magnetization injection and clearly dominates
it~\cite{LDAgFactor}.

Combining this with our calculations for the radiation emission
(following \cite{SipeJOSAB87}) we find that for GaAs the hole
radiation should be of the order $3\times 10^{-3}$, or about $1/300$
times the shift current radiation.

Despite the relatively large expected signal from the the holes, the
ratio of THz emission from magnetization to shift current is still
small, and one must choose an experimental geometry which minimizes
the shift current. Differences in the spatial symmetry
characteristics of shift current and magnetization sources are
exploited to distinguish the weak magnetization source from the much
stronger shift current source.

\begin{figure}
\includegraphics[scale=0.75]{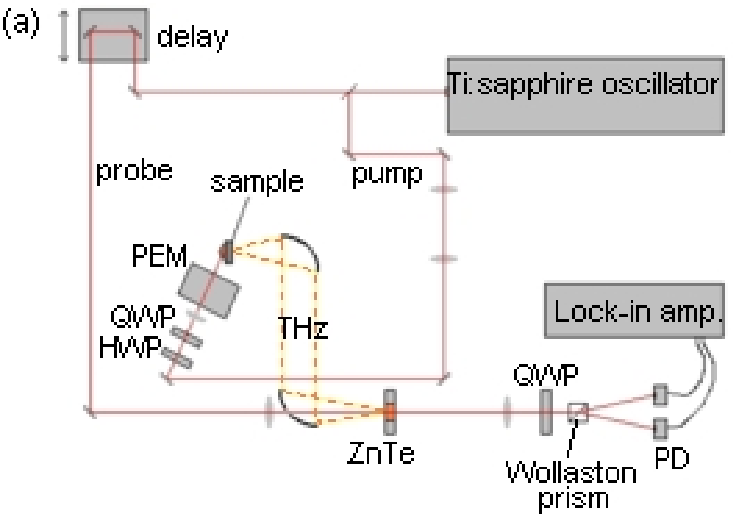} \\
\includegraphics[scale=0.75]{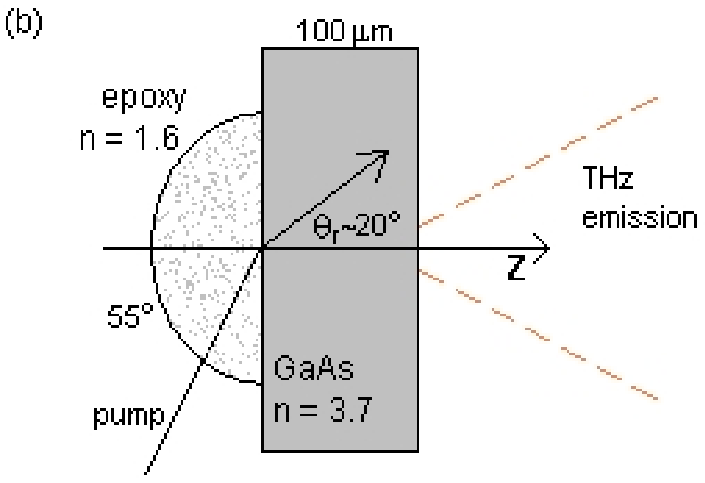} \\
\caption{(color online) (a) Experimental setup. (b) Top view of
sample with epoxy hemisphere.  PEM: photoelastic modulator, HWP:
half waveplate, QWP: quarter waveplate, PD: photodiode
\label{expsetup}}
\end{figure}

Fig. \ref{expsetup} illustrates the experimental setup for observing
the transient THz radiation.  An 80\,MHz Ti:sapphire oscillator
delivers 1\,nJ, 100\,fs pulses at 800\,nm.  For this wavelength, the
injected kinetic energy of heavy (light) holes is 3.6\,meV
(9.5\,meV), much less than the 300\,meV needed to excite holes in
the spin-orbit split-off band.  The pulses are split into a pump
beam for sample excitation and a probe beam to sample THz radiation
via electro-optic sampling.  The pump beam is incident onto a
100\,$\mu$m thick (110) GaAs sample with a spot diameter of
25\,$\mu$m, producing a peak incident intensity of 500\,MW/cm$^2$
and a carrier density of $2\times 10^{18}\,\text{cm}^{-3}$ over an
absorption depth of about 1\,$\mu$m. Since the total magnetization
or electric polarization is determined by total number of carriers
excited, and not their density, there is no premium associated with
beam focusing.  The emitted THz radiation emitted along the surface
normal direction from the sample back side is collected with a f =
50\,mm, $90^\circ$ off-axis parabolic and focused with the same type
of mirror onto a $500\,\mu$m thick (110) ZnTe electro-optic sampling
crystal. The $800\,\text{nm}$ probe pulse is spatially overlapped
with the THz field inside the ZnTe crystal and temporally scanned.
The pump beam polarization is controlled by half- and quarter-wave
plates (QWP) and a photo-elastic modulator (PEM). The half-wave
plate (HWP) is oriented to provide light linearly polarized
45$^\circ$ from the $\mathbf{\hat{s}}$ polarization vector (the
vertical). The PEM operates in quarter-wave mode and modulates the
polarization state at a frequency of 42\,kHz, enabling lock-in
detection techniques.

A major difficulty in measuring the emitted THz is that the injected
magnetic dipoles are parallel to the laser beam propagation
direction.  For normal incidence excitation, this injects magnetic
dipoles that are perpendicular to the surface.  The component of the
emitted radiation {nearly} parallel to the surface is small, and so
there is little transmission.  In the shift current effect however,
the polarization is parallel to the surface, and this allows the
radiated THz field to easily transmit through the surface.  To
enhance the emission of THz from the ultrafast optical orientation
it is thus necessary to irradiate the sample at {a high angle of
incidence; we use} an angle of incidence of $55^\circ$, exciting
through an epoxy hemisphere (see Fig. \ref{expsetup}b), to give an
angle of refraction inside the GaAs, $\theta_R$, of about
$20^\circ$. Even for this small angle it follows from the Maxwell
{\it saltus} (boundary) conditions that the emitted THz radiation
mainly propagates in the direction of the surface normal, because
the induced electron-hole plasma is $< 1\,\mu\text{m}$ thick,

We define $\beta$ as the angle between the QWP's fast axis and
$\mathbf{\hat{s}}$.  The shift current and magnetization signals
both depend on $\beta$ in a characteristic and unique way, so we
focus on that variable to distinguish between the two THz sources.
If we define $\gamma$ as the angle between the [001] crystal axis
(lying in the plane of the sample) and $\mathbf{\hat{s}}$, the
internal THz field emitted by the two types of sources are given by
\begin{eqnarray}
\Delta E_{\text{THz}}^{\text{shift}} & = & C e d \cos{(\theta_R)}
(2 \sin{\gamma} - 3 \sin^3{\gamma}) \cos{(2 \beta)}, \\
\Delta E_{\text{THz}}^{\text{mag}} & = & C \mu_{\text{eff}}
\sin{(\theta_R)} \sin^2{(2 \beta)},
\end{eqnarray}
where the proportionality constant $C$ includes universal constants
and Fresnel reflection and transmission coefficients.  The effective
magnetic moment $\mu_\text{eff}$ describes the sum of electron and
hole magnetic moments, although we expect it to be dominated by hole
term.  It is evident that the shift current (magnetization) signal
is maximized (minimized) when $\beta = 0^\circ$ and minimized
(maximized) when $\beta = 45^\circ$. This is because when
$\cos(2\beta)=0$ ($\sin(2\beta)=0$), the light polarization is
modulating between $+45^\circ$ and $-45^\circ$ linear (right and
left circular) states.  We choose $\gamma = 0.02\,\text{rad}$
($1.2^\circ$), for which $2\sin\gamma -3\sin^3\gamma\approx
2\gamma$, in order to provide a small shift current source with
which to compare the magnetization signal.

\begin{figure}
\includegraphics[scale=0.5]{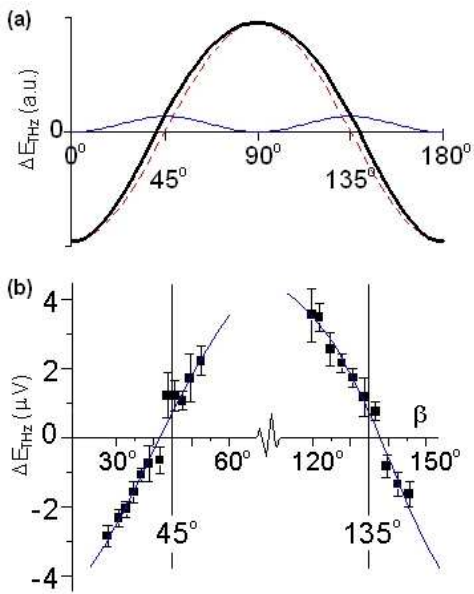}
\caption{(color online) (a) Simulation of THz signal vs QWP angle
$\beta$. red dashed: shift current signal, blue thin: magnetization
signal, black thick: total signal (b) Experimental results (points)
with Fourier series fit (curve) \label{expresults}}
\end{figure}

Fig. \ref{expresults}a illustrates the expected variation of the THz
field strength from the shift and magnetization sources, taking the
ratio of the source strengths to be 0.015.  Other possible sources
of THz radiation were neglected.  Optical rectification is small
relative to the shift current effect above the band gap, and shares
the same symmetry dependence.  Sources of THz radiation related to
Dember fields have no dependence on crystal orientation or pump
polarization.  Fig. \ref{expresults}b shows results from
experimental measurements of the THz field strength as a function of
$\beta$. The THz field strength is defined as the peak-to-peak value
of the observed THz trace, although several other definitions of
signal strength were used and each resulted in a similar pattern as
a function of $\beta$.

The deviation of the expected sinusoidal pattern created by the
shift current alone is attributed to the signal from the
magnetization source.  The "zero crossings" of the Fourier series
fit to the data of Fig. \ref{expresults}b are separated by more than
$90^\circ$. This behavior indicates the presence of an additional
THz source with a $\beta$-dependence as shown in Fig.
\ref{expresults}a. The difference between maximum and minimum signal
values recorded at $\beta=90^\circ$ and $0^\circ$ respectively was
$0.05\pm 0.14\,\mu$V. This difference would have had to be on the
order of $1\,\mu\text{V}$ to attribute the data's deviation to a
uniform vertical shift. Experimental data indicates that the
relative strengths of the magnetization and shift current sources
have a ratio 0.017, which is within a factor of six of the
theoretical value of the hole contribution.  More work is required
to confirm that the offset in Fig.~\ref{expresults}b is mainly due
to the magnetization source. Given all the possible sources of
systematic error, we consider this extremely good agreement in the
first study of this effect.

Increasing the THz emission from the spin injection can be achieved
in a few ways: The THz generation depends on the injection rate of
spin polarized carriers, and so using a shorter pulse will give a
larger injection rate and increase the THz signal.  One could also
use materials with a larger $g$-factor. Materials with $g$-factors
near 50, such as InAs, have been reported \cite{RothPR59}. Finally,
some materials also allow for a higher degree of spin polarization
$S$ to be achieved; in strained GaAs or the II-VI wurtzite
materials, for example, optical orientation can give an almost 100\%
polarized electron spin population.

In conclusion, we have investigated the possibility of measuring THz
radiation from the ultrafast optical orientation of electrons and
holes. This THz radiation is due to a rapidly-varying magnetic
source, and we have shown that it is large enough to be detected
against the more usual THz signal from electric current sources.
Measurements of the THz radiation from spin-injection can complement
the wide range of experiments on spin polarization decay already
existing in the literature, providing a direct measurement of the
injected spin dynamics of both electrons and holes in
semiconductors.

We thank J. Prineas, C. Pryor, A. Bristow, M. Betz and J. Rioux for
useful discussions.  Preliminary experimental work by Q. H. Wang is
also acknowledged.  This work is supported by NSERC, Photonics
Research Ontario, and DARPA SpinS.

\bibliographystyle{apsrev}
\bibliography{THz_biblio}

\begin{thebibliography}{24}
\expandafter\ifx\csname natexlab\endcsname\relax\def\natexlab#1{#1}\fi
\expandafter\ifx\csname bibnamefont\endcsname\relax
  \def\bibnamefont#1{#1}\fi
\expandafter\ifx\csname bibfnamefont\endcsname\relax
  \def\bibfnamefont#1{#1}\fi
\expandafter\ifx\csname citenamefont\endcsname\relax
  \def\citenamefont#1{#1}\fi
\expandafter\ifx\csname url\endcsname\relax
  \def\url#1{\texttt{#1}}\fi
\expandafter\ifx\csname urlprefix\endcsname\relax\def\urlprefix{URL }\fi
\providecommand{\bibinfo}[2]{#2}
\providecommand{\eprint}[2][]{\url{#2}}

\bibitem[{Dya()}]{DyakanovPerel}
\bibinfo{note}{M. I. Dyakanov and V. I. Perel, in {\it Optical Orientation},
  edited by F. Meier and B. P. Zakharchenya (Elsevier, Amsterdam, 1984), Chap.
  2, pp. 11-71.}

\bibitem[{\citenamefont{{\v{Z}}uti{\'c}
  et~al.}(2004)\citenamefont{{\v{Z}}uti{\'c}, Fabian, and Sarma}}]{ZuticRMP04}
\bibinfo{author}{\bibfnamefont{I.}~\bibnamefont{{\v{Z}}uti{\'c}}},
  \bibinfo{author}{\bibfnamefont{J.}~\bibnamefont{Fabian}}, \bibnamefont{and}
  \bibinfo{author}{\bibfnamefont{S.~D.} \bibnamefont{Sarma}},
  \bibinfo{journal}{Rev. Mod. Phys.} \textbf{\bibinfo{volume}{76}},
  \bibinfo{pages}{323} (\bibinfo{year}{2004}).

\bibitem[{\citenamefont{Hilton and Tang}(2002)}]{HiltonPRL02}
\bibinfo{author}{\bibfnamefont{D.~J.} \bibnamefont{Hilton}} \bibnamefont{and}
  \bibinfo{author}{\bibfnamefont{C.~L.} \bibnamefont{Tang}},
  \bibinfo{journal}{Phys. Rev. Lett.} \textbf{\bibinfo{volume}{89}},
  \bibinfo{pages}{146601} (\bibinfo{year}{2002}).

\bibitem[{\citenamefont{Yu et~al.}(2005)\citenamefont{Yu, Krishnamurthy, van
  Schilfgaarde, and Newman}}]{YuPRB05}
\bibinfo{author}{\bibfnamefont{Z.~G.} \bibnamefont{Yu}},
  \bibinfo{author}{\bibfnamefont{S.}~\bibnamefont{Krishnamurthy}},
  \bibinfo{author}{\bibfnamefont{M.}~\bibnamefont{van Schilfgaarde}},
  \bibnamefont{and} \bibinfo{author}{\bibfnamefont{N.}~\bibnamefont{Newman}},
  \bibinfo{journal}{Phys. Rev. B} \textbf{\bibinfo{volume}{71}},
  \bibinfo{pages}{245312} (\bibinfo{year}{2005}).

\bibitem[{\citenamefont{Oestreich et~al.}(2002)\citenamefont{Oestreich, Bender,
  H{\"u}bner, H{\"a}gele, R{\"u}hle, Hartmann, Klar, Heimbrodt, Lampalzer, Volz
  et~al.}}]{OestrichSST17}
\bibinfo{author}{\bibfnamefont{M.}~\bibnamefont{Oestreich}},
  \bibinfo{author}{\bibfnamefont{M.}~\bibnamefont{Bender}},
  \bibinfo{author}{\bibfnamefont{J.}~\bibnamefont{H{\"u}bner}},
  \bibinfo{author}{\bibfnamefont{D.}~\bibnamefont{H{\"a}gele}},
  \bibinfo{author}{\bibfnamefont{W.~W.} \bibnamefont{R{\"u}hle}},
  \bibinfo{author}{\bibfnamefont{T.}~\bibnamefont{Hartmann}},
  \bibinfo{author}{\bibfnamefont{P.~J.} \bibnamefont{Klar}},
  \bibinfo{author}{\bibfnamefont{W.}~\bibnamefont{Heimbrodt}},
  \bibinfo{author}{\bibfnamefont{M.}~\bibnamefont{Lampalzer}},
  \bibinfo{author}{\bibfnamefont{K.}~\bibnamefont{Volz}}, \bibnamefont{et~al.},
  \bibinfo{journal}{Semicond. Sci. Technol.} \textbf{\bibinfo{volume}{17}},
  \bibinfo{pages}{285} (\bibinfo{year}{2002}).

\bibitem[{\citenamefont{Awschalom and Kikkawa}(1999)}]{AwschalomPT52}
\bibinfo{author}{\bibfnamefont{D.~D.} \bibnamefont{Awschalom}}
  \bibnamefont{and} \bibinfo{author}{\bibfnamefont{J.~M.}
  \bibnamefont{Kikkawa}}, \bibinfo{journal}{Phys. Today}
  \textbf{\bibinfo{volume}{52}}, \bibinfo{pages}{33} (\bibinfo{year}{1999}).

\bibitem[{\citenamefont{Beaurepaire et~al.}(2004)\citenamefont{Beaurepaire,
  Turner, Harrel, , Beard, Bigot, and Schmuttenmaer}}]{BeaurepaireAPL04}
\bibinfo{author}{\bibfnamefont{E.}~\bibnamefont{Beaurepaire}},
  \bibinfo{author}{\bibfnamefont{G.~M.} \bibnamefont{Turner}},
  \bibinfo{author}{\bibfnamefont{S.~M.} \bibnamefont{Harrel}}, ,
  \bibinfo{author}{\bibfnamefont{M.~C.} \bibnamefont{Beard}},
  \bibinfo{author}{\bibfnamefont{J.-Y.} \bibnamefont{Bigot}}, \bibnamefont{and}
  \bibinfo{author}{\bibfnamefont{C.~A.} \bibnamefont{Schmuttenmaer}},
  \bibinfo{journal}{Appl. Phys. Lett.} \textbf{\bibinfo{volume}{84}},
  \bibinfo{pages}{3465} (\bibinfo{year}{2004}).

\bibitem[{\citenamefont{Schmuttenmaer}(2004)}]{SchmuttenmaerCR04}
\bibinfo{author}{\bibfnamefont{C.~A.} \bibnamefont{Schmuttenmaer}},
  \bibinfo{journal}{Chem. Rev.} \textbf{\bibinfo{volume}{104}},
  \bibinfo{pages}{1759} (\bibinfo{year}{2004}).

\bibitem[{\citenamefont{Johnston et~al.}(2002)\citenamefont{Johnston,
  Whittaker, Corchia, Davies, and Linfield}}]{JohnstonPRB02}
\bibinfo{author}{\bibfnamefont{M.~B.} \bibnamefont{Johnston}},
  \bibinfo{author}{\bibfnamefont{D.~M.} \bibnamefont{Whittaker}},
  \bibinfo{author}{\bibfnamefont{A.}~\bibnamefont{Corchia}},
  \bibinfo{author}{\bibfnamefont{A.~G.} \bibnamefont{Davies}},
  \bibnamefont{and} \bibinfo{author}{\bibfnamefont{E.~H.}
  \bibnamefont{Linfield}}, \bibinfo{journal}{Phys. Rev. B}
  \textbf{\bibinfo{volume}{65}}, \bibinfo{pages}{165301}
  (\bibinfo{year}{2002}).

\bibitem[{\citenamefont{C\^ot\'e et~al.}(2002)\citenamefont{C\^ot\'e, Laman,
  and van Driel}}]{CoteAPL02}
\bibinfo{author}{\bibfnamefont{D.}~\bibnamefont{C\^ot\'e}},
  \bibinfo{author}{\bibfnamefont{N.}~\bibnamefont{Laman}}, \bibnamefont{and}
  \bibinfo{author}{\bibfnamefont{H.~M.} \bibnamefont{van Driel}},
  \bibinfo{journal}{Appl. Phys. Lett.} \textbf{\bibinfo{volume}{80}},
  \bibinfo{pages}{905} (\bibinfo{year}{2002}).

\bibitem[{\citenamefont{von Baltz and Kraut}(1981)}]{vonBaltzPRB81}
\bibinfo{author}{\bibfnamefont{R.}~\bibnamefont{von Baltz}} \bibnamefont{and}
  \bibinfo{author}{\bibfnamefont{W.}~\bibnamefont{Kraut}},
  \bibinfo{journal}{Phys. Rev. B} \textbf{\bibinfo{volume}{23}},
  \bibinfo{pages}{5590} (\bibinfo{year}{1981}).

\bibitem[{not()}]{noteAboutPGEffect}
\bibinfo{note}{There are configurations for which the photogalvanic effect
  appears with circular polarized light. However, these configurations are
  avoided in this work and for our discussion it can be assumed that the
  photogalvanic effect is due to linearly polarized excitation.}

\bibitem[{\citenamefont{Nastos and Sipe}(2006)}]{NastosPRB06}
\bibinfo{author}{\bibfnamefont{F.}~\bibnamefont{Nastos}} \bibnamefont{and}
  \bibinfo{author}{\bibfnamefont{J.~E.} \bibnamefont{Sipe}},
  \bibinfo{journal}{Phys. Rev. B} \textbf{\bibinfo{volume}{74}},
  \bibinfo{pages}{035201} (\bibinfo{year}{2006}).

\bibitem[{neg()}]{neglectOR}
\bibinfo{note}{In addition to the shift current, there is also an optical
  rectification rectification current which can radiate THz for the same
  experimental configurations. However, for above band gap excitation the shift
  current dominates in GaAs~\cite{CoteAPL02}.}

\bibitem[{\citenamefont{Roth}(1960)}]{RothPR60}
\bibinfo{author}{\bibfnamefont{L.~M.} \bibnamefont{Roth}},
  \bibinfo{journal}{Phys. Rev.} \textbf{\bibinfo{volume}{118}},
  \bibinfo{pages}{1534} (\bibinfo{year}{1960}).

\bibitem[{\citenamefont{Chadi et~al.}(1976)\citenamefont{Chadi, Clark, and
  Burnham}}]{ChadiPRB76}
\bibinfo{author}{\bibfnamefont{D.~J.} \bibnamefont{Chadi}},
  \bibinfo{author}{\bibfnamefont{A.~H.} \bibnamefont{Clark}}, \bibnamefont{and}
  \bibinfo{author}{\bibfnamefont{R.~D.} \bibnamefont{Burnham}},
  \bibinfo{journal}{Phys. Rev. B} \textbf{\bibinfo{volume}{13}},
  \bibinfo{pages}{4466} (\bibinfo{year}{1976}).

\bibitem[{\citenamefont{Hermann and Weisbuch}(1977)}]{HermannPRB77}
\bibinfo{author}{\bibfnamefont{C.}~\bibnamefont{Hermann}} \bibnamefont{and}
  \bibinfo{author}{\bibfnamefont{C.}~\bibnamefont{Weisbuch}},
  \bibinfo{journal}{Phys. Rev. B} \textbf{\bibinfo{volume}{15}},
  \bibinfo{pages}{823} (\bibinfo{year}{1977}).

\bibitem[{\citenamefont{Lloyd-Hughes et~al.}(2004)\citenamefont{Lloyd-Hughes,
  Castro-Camus, Fraser, Jagadish, and Johnston}}]{LloydHughesPRB04}
\bibinfo{author}{\bibfnamefont{J.}~\bibnamefont{Lloyd-Hughes}},
  \bibinfo{author}{\bibfnamefont{E.}~\bibnamefont{Castro-Camus}},
  \bibinfo{author}{\bibfnamefont{M.~D.} \bibnamefont{Fraser}},
  \bibinfo{author}{\bibfnamefont{C.}~\bibnamefont{Jagadish}}, \bibnamefont{and}
  \bibinfo{author}{\bibfnamefont{M.~B.} \bibnamefont{Johnston}},
  \bibinfo{journal}{Phys. Rev. B} \textbf{\bibinfo{volume}{70}},
  \bibinfo{pages}{235330} (\bibinfo{year}{2004}).

\bibitem[{\citenamefont{Sipe}(1987)}]{SipeJOSAB87}
\bibinfo{author}{\bibfnamefont{J.~E.} \bibnamefont{Sipe}}, \bibinfo{journal}{J.
  Opt. Soc. Am. B} \textbf{\bibinfo{volume}{4}}, \bibinfo{pages}{481}
  (\bibinfo{year}{1987}).

\bibitem[{\citenamefont{Pryor and Flatte}(2006)}]{PryorPRL06}
\bibinfo{author}{\bibfnamefont{C.~E.} \bibnamefont{Pryor}} \bibnamefont{and}
  \bibinfo{author}{\bibfnamefont{M.~E.} \bibnamefont{Flatte}},
  \bibinfo{journal}{Phys. Rev. Lett.} \textbf{\bibinfo{volume}{96}},
  \bibinfo{pages}{026804} (\bibinfo{year}{2006}).

\bibitem[{\citenamefont{Richard et~al.}(2004)\citenamefont{Richard, Aniel, and
  Fishman}}]{RichardPRB04}
\bibinfo{author}{\bibfnamefont{S.}~\bibnamefont{Richard}},
  \bibinfo{author}{\bibfnamefont{F.}~\bibnamefont{Aniel}}, \bibnamefont{and}
  \bibinfo{author}{\bibfnamefont{G.}~\bibnamefont{Fishman}},
  \bibinfo{journal}{Phys. Rev. B} \textbf{\bibinfo{volume}{70}},
  \bibinfo{pages}{235204} (\bibinfo{year}{2004}).

\bibitem[{\citenamefont{Bhat et~al.}(2005)\citenamefont{Bhat, Nastos, Najmaie,
  and Sipe}}]{BhatPRL05}
\bibinfo{author}{\bibfnamefont{R.~D.~R.} \bibnamefont{Bhat}},
  \bibinfo{author}{\bibfnamefont{F.}~\bibnamefont{Nastos}},
  \bibinfo{author}{\bibfnamefont{A.}~\bibnamefont{Najmaie}}, \bibnamefont{and}
  \bibinfo{author}{\bibfnamefont{J.~E.} \bibnamefont{Sipe}},
  \bibinfo{journal}{Phys. Rev. Lett.} \textbf{\bibinfo{volume}{94}},
  \bibinfo{pages}{096603} (\bibinfo{year}{2005}).

\bibitem[{LDA()}]{LDAgFactor}
\bibinfo{note}{This large $\tilde{g}_h$ value was verified with an LDA
  pseudopotential calculation, where we found $\tilde{g}_h = 19$. Even though
  there is some disagreement between the two methods, they both point to a
  $\tilde{g}_h$ that is at least 30 times larger than $g_e$.}

\bibitem[{\citenamefont{Roth et~al.}(1959)\citenamefont{Roth, Lax, and
  Zwerdling}}]{RothPR59}
\bibinfo{author}{\bibfnamefont{L.~M.} \bibnamefont{Roth}},
  \bibinfo{author}{\bibfnamefont{B.}~\bibnamefont{Lax}}, \bibnamefont{and}
  \bibinfo{author}{\bibfnamefont{S.}~\bibnamefont{Zwerdling}},
  \bibinfo{journal}{Phys. Rev.} \textbf{\bibinfo{volume}{114}},
  \bibinfo{pages}{90} (\bibinfo{year}{1959}).

\end{thebibliography}

\end{document}